\documentstyle[12pt]{article}
\begin{document}

\begin{center} {\bf \Huge Condensation and Evaporation of Mutually Repelling Particles
  :Steady states and limit cycles}
 \vskip .5cm Tapati Dutta$^1$,  Nikolai Lebovka$^2$ and S.
Tarafdar$^3$\\ $^1$ Physics Department, St. Xavier's College,\\
Kolkata 700016, India\\ $^2$F. D. Ovcharenko Biocolloid Chemistry Institute \\42,
Vernadsky Av., Kyiv, Ukraine\\ $^3$ Condensed Matter Physics Research Centre,\\Physics
Department, Jadavpur University,\\ Kolkata 700032, India.\\Email:
sujata@juphys.ernet.in (S Tarafdar)\\
\end{center}

\noindent {\bf Abstract}

We study condensation and evaporation of particles which repel each other, using a
simple set of rules on a square lattice. Different results are obtained for a mobile
and an immobile surface layer.A two point limit cycle is observed for high temperature
and low pressure in both cases. Here the coverage oscillates between a high and a low
value without ever reaching a steady state. The results for the immobile case depend
in addition on the initial coverage.

\section{Introduction} Recently computer
simulations on cellular automata type models \cite{ce,manna,ed}have been suggested
which can be applied to simulate the development of a surface layer under different
physical conditions. A randomly occupied square lattice is taken and particles are
allowed to stick at vacant sites or evaporate from occupied sites according to a
prescribed set of rules. Appropriate probabilities for the `growing' and `culling' are
set according to temperature and pressure conditions required.We can also introduce an
effective interaction between the particles by making the probabilities a function of
the surroundings of a site, e.g. the number of nearest neighbours.

In our earlier work \cite{ce} we started with a randomly occupied square lattice and
let particles evaporate with a probability which decreased with the number of occupied
neighbouring sites. This amounted to an attractive interaction between the particles.
The condensation was a function of only the superincumbent pressure. The present paper
is an extension of the earlier work, here we introduce an effective repulsive
interaction between particles, by allowing the condensation probability at an empty
site to decrease with the number of occupied neighbours. The evaporation probability
is a function of temperature only. The results are very interesting. We study as in
the previous work \cite{ce}  evolution of a mobile layer by analytical and numerical
methods and an immobile layer by computer simulation. In both cases we find a steady
final coverage under some conditions, but a two-point limit cycle showing oscillations
in the coverage when temperature is high and pressure low.

In the next section we describe the model. The results are presented in the third
section and discussion and conclusions in the last.

\section{The Model} The process of condensation/evaporation or adsorption/desorption
is described
traditionally by two different sets of models -- one for a mobile adsorption layer and
one for an immobile layer\cite{book1,book2}. Probabilities for sticking and
evaporation on a two-dimensional monolayer are specified according to the physics
behind the model.These are functions of the temperature, superincumbent pressure and
the existing coverage. At equilibrium, the sticking and evaporation probabilities are
set equal and the resulting equation is solved to get the equilibrium coverage at that
temperature and pressure.

The simplest mobile layer model is a two-dimensional ideal gas, and the improved
versions include interaction between particles, similar to a two-dimensional Van der
Waal gas. The so-called immobile layer models introduce a sticking probability,
depending on how long a molecule in the vapor above the surface is in contact with a
surface site. The simplest `immobile model' is the
 Langmuir equation derived as follows.

For vapor condensation the particle flux, i.e. the number of particles deposited per
unit time per unit surface area  is equal to:
\begin{equation}
c=\frac{P\lambda}{h} \label{e1}
\end{equation}
where
 $P$ is the pressure, $$\lambda=\sqrt{h^{2}/(2\pi mkT)}$$ is de
Broglie length, $h$ is Planck constant.
The probability of condensation per unit time is obtained by dividing $c$ by the total number
of sites available for condensation. And the same may be done for the evaporation probability.
Flux of evaporation  from saturated surface (at $p=1$) may be approximated by
\cite{jay} :
\begin{equation}
d=\frac{kT}{h\lambda^{2}}\exp(-E_{e}/kT) \label{e2}
\end{equation}

Condensation probability is set equal to evaporation probability, giving equilibrium:

\begin{equation}
\frac{P\lambda^{3}}{kT}(1-p)=\exp(-E_{e}/kT)p \label{e4}
\end{equation}
and we have the simple Langmuir equation
\begin{equation}
Pb=\frac{p}{1-p} \label{e5}
\end{equation}
where
$$b=\frac{\lambda^{3}}{kT}\exp(E_{e}/kT),$$
$E_e$ is activation energy for evaporation.

In this approximation it is assumed that evaporation energy is the same for  all
configurations.

In the present work we introduce a repulsive interaction between particles, as follows
- condensing particles are less likely to stick at sites surrounded by more neighbors.
This mimics a situation where the adsorbing particle has to overcome the potential
barrier due the repelling neighbors to reach a vacant site on the substrate. It is
implied that the substrate attracts the adsorbing particle.

 We define  a  probability for condensation on a vacant site $f_c$ determined by the pressure  as
\begin{equation} f_c = P/P_s \label{e6} \end{equation}
where $P_s$ is a saturation pressure and $0<f_c<1$.
The growth probability at a vacant site with $n$ occupied neighbors is proportional to
${f_c}^n$.  We have
\begin{equation} P_{gr} = (1-p)^5 {f_c}^4 + 4(1-p)^4 p {f_c}^3 +
6(1-p)^3p^2 {f_c}^2 + 4(1-p)^2p^3 {f_c} + (1-p)p^4 \label{imp}
\end{equation}
The successive terms in the RHS above represent the probability of a vacant site
having respectively 4,3,2,1 and 0 vacant near neighbor sites. $P_{gr}$ can be written
in the compact form
\begin{equation} P_{gr} = (1-p)(p{f_c}+1-p)^4 \label{e8} \end{equation}

The culling probability is assumed to be a function of temperature only
\begin{equation} P_{cull} =p(kT/ {P_s{\lambda}^3)} exp(-E_e/kT)
\label{e9} \end{equation}

$P_s$ is temperature dependent, and this must be taken into account. But we can approximate it by the simple relation that follows from
equilibrium relation on the surface of fluid
\begin{equation}
P_s=\frac{kT}{\lambda^{3}}\exp(-E_{vap}/kT)
 \end{equation}
where $E_{vap}$ is the enthalpy of vaporization of a fluid that
supports vapour.

In this formulation the  simplest condensation evaporation
equilibrium condition is

\begin{equation}
\frac{P}{P_s}(1-p)=\frac{kT}{P_s\lambda^{3}}\exp(-E_{e}/kT)p=\exp{-\Delta
E/kT}p  \end{equation}

where $\Delta E= E_{e}-E_{vap}$

Here too condensation probability $f_c$ (or reduced pressure $P^*$) is
\begin{equation}
f_c=P/P_s
\end{equation} and the evaporation probability is
\begin{equation}
f_e=\exp(-\Delta E/kT)
\end{equation}

That means that the molecules evaporate from the our surface with
probability 1 when $E_{e}=E_{vap}$, when energy of evaporation
from surface  $E_{e}$ equals to energy of evaporation from surface
of liquid that support vapor $E_{vap}$.

So, we can rewrite
$ P_{cull}$ as

\begin{equation} P_{cull} =p\exp(-1/T^*)
 \end{equation}
where $f_c=P/P_s$ and $T^*=kT/\Delta E$

Our model is complementary to the Fowler-Guggenheim model \cite{book1,book2}, where
the condensation probability is dependent only on pressure, but the evaporation
probability decreases with the number of occupied near neighbors. The neighbors are
assumed to attract the particle and create a barrier to evaporation. This type of
interaction was incorporated in Dutta  et al \cite{ce}.

Now we study the behavior of this system for different values of the temperature and
pressure parameters. We look at the two different situations the mobile surface layer
and the immobile layer.
\subsection{Mobile surface layer}
Let us suppose that the energy barrier for motion of the deposited particles along the
surface is very low so a continuous rearrangement is going on. In this case the
distribution of particles always becomes random and the probabilities in eq(\ref{imp})
are always valid. So one can equate the growth and culling probabilities in the steady
state and solve for the coverage $p$. We confine ourselves to ranges of $T^{\star}$
and $P^*$, which give growth and culling probabilities in the range 0 -- 1. We find that
meaningful positive values for the coverage between 0 and 1 are obtained for a very
narrow range of temperature and pressure. The results are shown in the next section.
We may also see the how the coverage evolves using the following iterative procedure
\begin{equation}
p_f = p_i + P_{gr} -P_{cl} 
\end{equation}
and
\begin{equation}
p_i = p_f 
\end{equation}
here $p_i$ is the initial coverage and $p_f$ the final. If there is a steady solution
the final coverage $p_f$ converges to a steady value. We find however that in some
combinations of the pressure and temperature parameters the coverage converges to a
stable limit cycle, oscillating between two positive and realistic values. The solution
for p in eq.(\ref{imp}) now becomes an unstable fixed point. This is very interesting and
reminiscent of the discovery of oscillatory behaviour in real chemical systems
\cite{chaos,osc1,osc2}. 

The results obtained here can be verified from a computer simulation on a finite
system. This was done as follows.

In the mobile case, the adsorbed molecules can move laterally on the surface.  The
physical situation simulated is described by eq.(\ref{imp}) . A two-dimensional square
lattice of unit spacing and size $300 \times 300$, is occupied randomly with an
initial coverage $p_{initial}$. Every occupied site is assigned the value $1$, empty
sites are assigned the value $0$. The occupied sites are then culled parallally with a
probability which is the same for each occupied site
$$g=exp({-1/T^*})$$ . For very large $T^*$ the culling probability approaches 1.
The vacant sites are  filled with a
probability dependent on the number of vacant near neighbor sites n as ${f_c}^n$.
 After one round of growth and culling is
complete, the concentration of the occupied sites $p_{final}$ is calculated.

In the next time-step, the $p_{final}$ of the previous time-step becomes the new
$p_{initial}$. The square lattice is then randomly occupied afresh with this
$p_{initial}$. A complete time-step begins with the random occupation of all sites
with a $p_{initial}$ and ends with the assignment of the $p_{final}$ to the
$p_{initial}$ of the next time-step. This iterative process stops when  $p_{final}$
saturates with increasing time to a definite value. In the simulation, we checked up
to 50,000 time-steps. The `mobility' of the molecules is simulated by the
randomization of the concentration $p_{initial}$ in the beginning of every time-step.

\subsection{Immobile Surface Layer}

If we consider the energy barrier to surface movement of the particles to be  large
enough the  arguments  presented in the last subsection are not valid. We start with a
certain concentration of particles randomly distributed on a square lattice and allow
growing and culling according to our rules. But after one time-step the distribution
is no longer random, so we cannot find out the number of new growth or evaporation
sites as in eq.(\ref{imp}). We can however simulate the system as above without the
randomization procedure after each round of growth and culling. This mimics the
development of a surface layer where the particles are not free to move laterally on
the surface. The results in this case are quite different .

\section{Results}
We present the results for the mobile and immobile layer case. The mobile layer
results are done analytically or by numerical iteration and verified by computer
simulation. The immobile layer results are done by computer simulation as calculation
is not possible here.
 \subsection{Results for the mobile layer}
 The results are shown in figure 1. The final coverage $p$ for each $P^*$, on 
 increasing $T^*$, from a low value
shows a bifurcation at a certain point to a two state limit cycle. The temperature and coverage at this point are
denoted respectively by ${T^*}_b$ and $p_b$, the inset shows the variation of ${T^*}_b$ and $p_b$ with $P^*$.
 
 These results are independent of the initial coverage.
For $T^*$ very large and $P$=1  there is an oscillation between $p_{initial}$ and $1-p_{initial}$, as is also evident from
eq.(\ref{imp}).
\subsection{Results for the immobile surface layer}

The immobile situation can be studied only by computer simulation. Starting with an
initial random distribution of particles the system is allowed to evolve and the final
coverage for different $P$ and $T^*$ is studied. In this case as there is no
randomization, the system takes a very long time to reach the steady state, and we
have observed up to 30,000 time steps.
Steady state solutions of $p_{final}$ are obtained for $P_{cull} < 1$. The steady state isobars
and isotherms are shown in Fig.2. and Fig.3. respectively. At $P_{cull}$=1.0, oscillations between a 
two-point limit sets in for all values of $P$. However for every $P$, there is a certain initial coverage, $p_{\alpha}$
that yields a final coverage oscillating in a narrower confine between $0.4$ and $0.5$
 for all $P$. Even a departure of 0.05 from $p_{\alpha}$, shifts the final coverage to the fixed point limits of 
the wider limit cycle.  The
 rate of convergence to $p_{final}$ from different $p_{initial}$s
other than $p_{\alpha}$, is slower for higher values of $P$. For $p_{initial}$ very close to $p_{\alpha}$
in the case of $P=0.9$, we checked for 50,000 time steps to detect whether the final coverage 
was approaching the fixed points of the wider limit cycle. The $p_{\alpha}$ increases slowly with 
$P$ as displayed in Table.I. 		 
\section{Discussion}

We have demonstrated that the condensation/evaporation problem seen as a modification
of the percolation problem on a square lattice shows very interesting possibilities.
There is firstly, the question of mobile and immobile layer models of chemical physics
\cite{book1}. From our point of view equating the growth and culling probabilities to
find the steady state coverage is valid only when the particles can rearrange very
fast to preserve the random distribution. And we demonstrate that making the particles
truly immobile,on the two-dimensional surface gives a quite different final coverage.
This is also dependent on the initial coverage. This was discussed in the previous
work \cite{ce} with an attractive interaction between particles. Careful experiments
are needed to test the justification of these arguments.

The most striking result of this work is the observation of oscillations in the
condensation/evaporation problem, using such an extremely simple model, without taking any details of the
system into account. Earlier well known work on non-linear mathematical
equations \cite{chaos} show the appearance of bifurcations, repeated period doubling
sometimes leading to chaos. And such situations are found in real life in problems of
physics, chemistry and biology. Chemical systems are known to exhibit oscillations in
concentration of reactants and products in chemical reactions \cite{osc1,osc2}. There is also some study of such a
 phenomenon  in condensation/evaporation problems \cite{pat,zhd}.
Experiments with repelling particles adsorbing on a substrate are reported \cite{rep},
but in this experiment two different types of particles are involved.

One possibility of having a repulsive interaction is with similarly charged particles
condensing on an oppositely charged substrate.

So in conclusion the simple stochastic model with growth and evaporation on a square lattice gives rise to a host of interesting phenomena. With an attractive interaction between particles phase transitions and hysteresis was observed \cite{ce}, 
While a repulsive interaction gives oscillations in coverage under certain conditions.

\begin{tabular}{|c|c|c|c|c|c|c|c|c|} \hline

$P$& 0.1 & 0.2 & 0.3 & 0.5 & 0.6 & 0.7 & 0.8 & 0.9 \\ \hline
$p_{\alpha}$& 0.23 & 0.25 & 0.265 & 0.305 & 0.33 & 0.35 & 0.39 & 0.43 \\ \hline
\end{tabular} \vskip .5 cm
\vskip .5cm \noindent Table I : The variation of $p_{\alpha}$ with $P$
\vskip .5cm 

\noindent {\bf Figure captions}:
\vskip .5cm \noindent
Figure 1 : Isobars for the final coverage for the mobile layer. At ${T^*}_b$, the steady solution for $p_b$ 
shows the bifurcation to a limit cycle. The inset shows variation of ${T^*}_b$ and $p_b$ with $P^*$.
\vskip .5cm \noindent
Figure 2 : Isobars for the immobile case showing variation of coverage with 
$exp(-1/T^*)$. The values of $P$ for each curve are shown in the figure.
\vskip .5cm \noindent
Figure 3 : Isotherms for the immobile case showing variation of coverage with pressure $P$. 
Stable $p$ are obtained for $exp(-1/T^*)$ upto 0.9, for $exp(-1/T^*)$=1.0 oscillations are obtained.

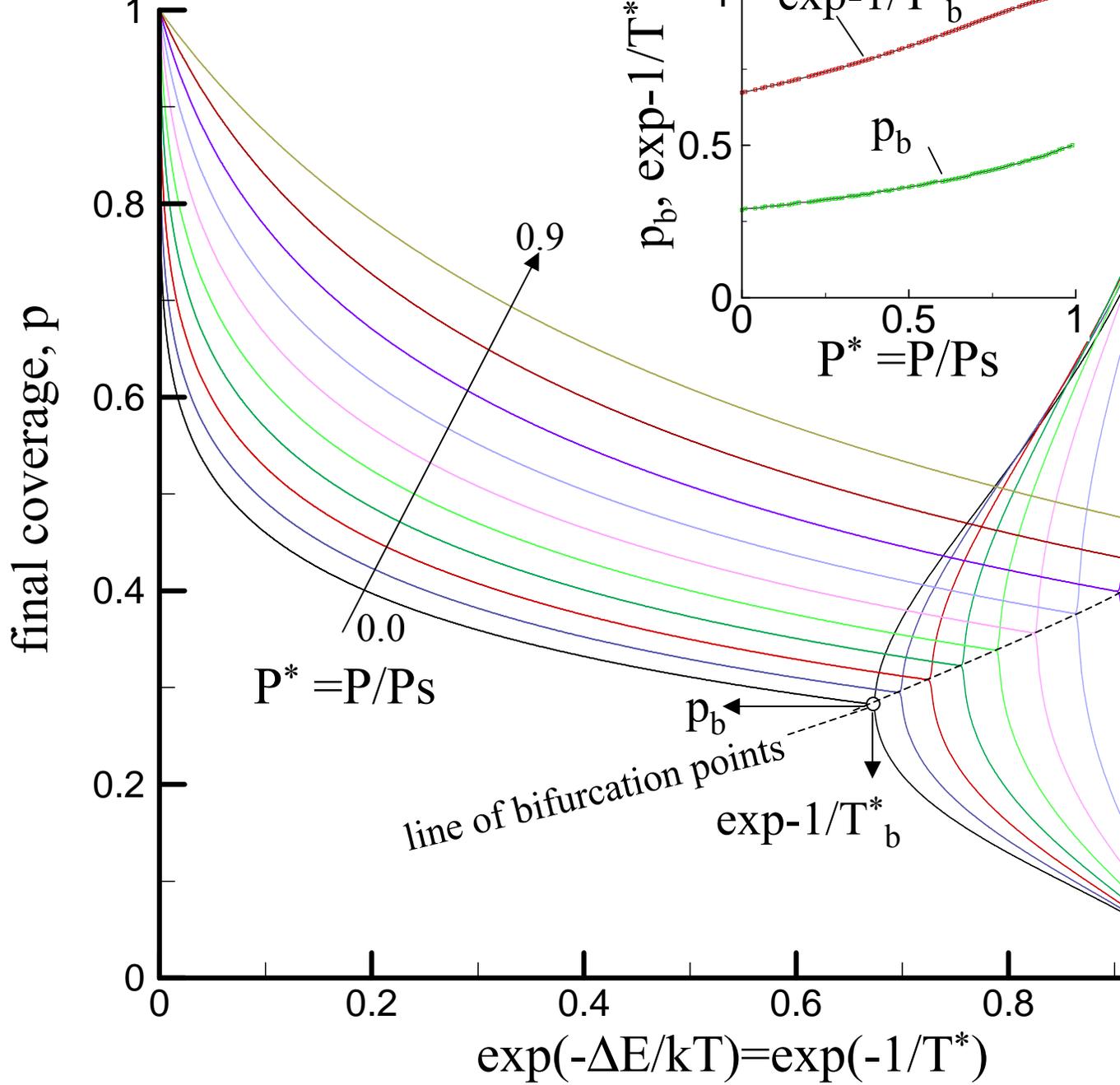
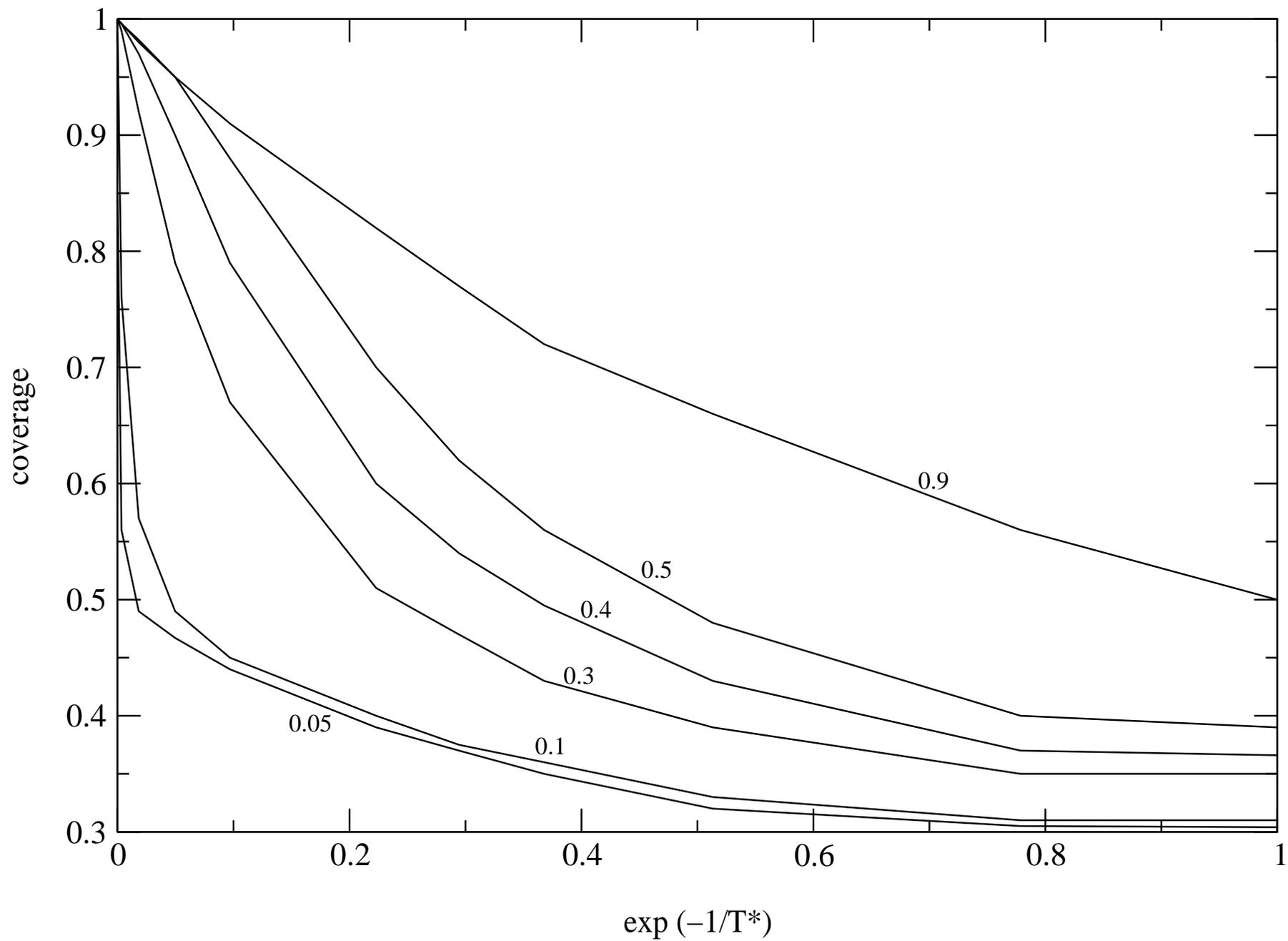
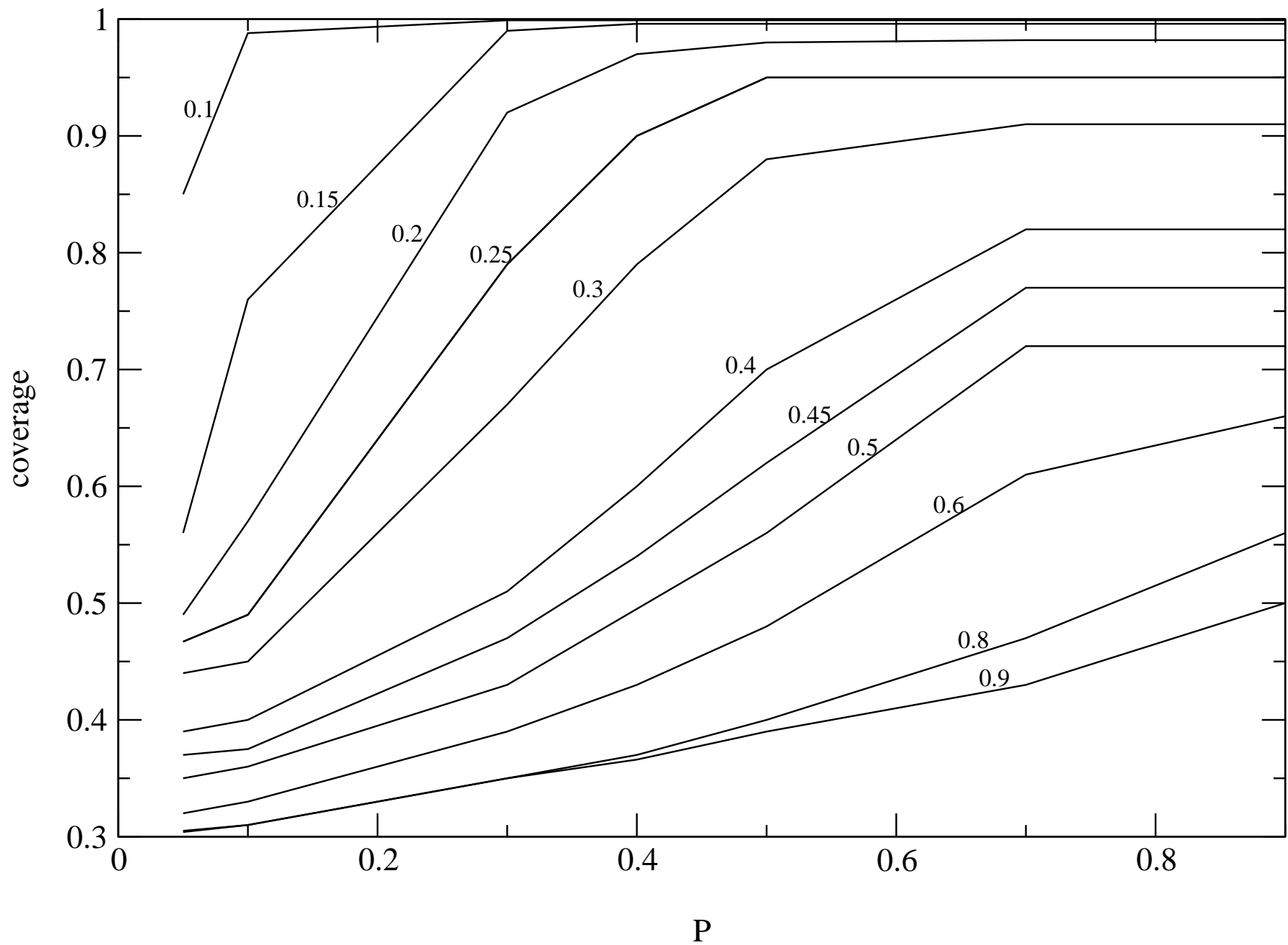
\end{document}